# Dependence of single molecule junction conductance on molecular conformation

Latha Venkataraman[1,4], Jennifer E. Klare[2,4], Colin Nuckolls[2,4], Mark S. Hybertsen[3,4], Michael L. Steigerwald[2]

*[1]Department of Physics, [2]Department of Chemistry, [3]Department of Applied Physics and Applied Mathematics, and [4]Center for Electron Transport in Molecular Nanostructures, Columbia University, New York, New York*

**The conductance of a single metal-molecule-metal junction depends not only on the chemical nature of the molecule used, but also on its conformation[1-4]. In the simple case of a biphenyl—two phenyl rings linked by a single C-C bond—conductance is expected to change with the relative twist angle between the two rings, with the planar conformation having the highest conductance. A number of different techniques have measured the conductance of metal-molecule(s)-metal junctions[5-12]. However, the conductance variation from junction to junction has made it difficult to verify even the simplest predictions about how molecules should behave in unimolecular devices. Here, using amine link groups[13] to form single molecule junctions, we show a clear correlation between molecule conformation and junction conductance in a series of seven biphenyl molecules with different ring substitutions that alter the twist angle of the molecules. We find that the conductance for the series decreases with increasing twist angle, consistent with a cosine squared relation predicted theoretically for transport through π-conjugated systems[14].**

We recently demonstrated that metal-molecule-metal junctions, formed by breaking Au point contacts in a solution of molecules, exhibit more reliable and reproducible conductance values when amine groups rather than thiols or isonitriles are used to attach the molecules to the junction contacts[13]. Because of this reduced variability, we can determine statistically meaningful average conductance values for specific single-molecule junctions; this capability in turn allows us to study the impact of molecular properties on junction conductance.

We present our experimental results as conductance histograms, where peaks indicate the most prevalent molecular junction conductances while the width of the conductance distributions reflects the microscopic variations from junction to junction. (For details on experimental and data analysis procedures, see Supplementary Information.) Figure 1A shows the histograms for 1,4-diaminobenzene (**1**) and 2,7-diaminofluorene (**2**), each constructed from over 10000 conductance traces without resorting to any data selection or processing. Many of our conductance traces reveal step-wise changes in conductance not only at conductance values that are multiples of the fundamental quantum of conductance $G_0$ ($2e^2/h$), but also below $G_0$ (see Fig. 1C and Supplementary Information Figure S1). These steps are due to conduction through a single molecule bridging the gap between the two Au point-contacts. As seen in the histograms of **1** and **2** (Fig. 1A), for each junction type the additional step occurs within a narrow distribution of conductance values. The most prevalent value is determined by a fit to the peak below $G_0$ in the conductance histograms using a Lorentzian line shape, which we find to fit our peaks more accurately than a Gaussian line shape (see inset of Figure 1A and also Supplementary Information).

When the experiment is repeated using a solution containing an equimolar mixture of **1** and **2** (as indicated in Figure 1B), the resulting histogram (red curve in Figure 1A) shows two distinct peaks below $G_0$ at nearly the same conductance values as those of the peaks seen in the histograms for the individual molecules (green and yellow curves). Individual traces using the solution mixture show conductance plateaus corresponding to either molecule **1** or **2**, and sometimes a plateau corresponding to **1** followed by a plateau corresponding to **2** (see Figure 1C). But we never see evidence of rapid fluctuations between the two, so the molecules are not exchanging position within the junction on the timescale of the experimental traces (few milliseconds for each trace). This provides further evidence that the conduction is through an individual molecule, not an ensemble.

Molecules **1** and **2** do not have an internal twist degree of freedom, but the two benzene rings of a biphenyl can rotate relative to one another. For this molecule, as the twist angle between the two rings increases and simultaneously the degree of π-conjugation between them decreases, the junction conductance will decrease because molecular electron transfer rates scale as the square of the π-overlap[15]. When neglecting the contribution of tunnelling transport through the σ-orbitals (which is within the error of our measurements), then theory predicts a $\cos^2\theta$ relation[14]. We probe this relation between molecule twist angle and conductance using seven different biphenyl molecules (molecules **2** to **8** listed in Table 1), where the size of the substituents on the proximal carbons is optimized to create structures with a twist angle ranging from flat to essentially perpendicular. Figure 2A shows the structure of four out of the seven molecules (**2**, **4**, **6**, and **8**) used. For each molecule, the average junction conductance is determined from the location of the peak in the conductance histogram, while the twist angle is determined theoretically for each molecule by assuming a fully relaxed and static geometry (see Table 1 and Supplementary Information; but effects of dynamic structural fluctuations of the molecule in the junction will also be discussed later on).

Figure 2B shows the measured conductance histograms for the molecules in Figure 2A. For the flat molecule (**2**), the histogram peak yields an average junction conductance value of $1.5\times10^{-3}$ $G_0$. For 4,4'-diaminobiphenyl (**4**) with equal molecular length but a twist angle θ of 34°, the conductance histogram shows a peak occurring at a lower conductance of $1.1\times10^{-3}$ $G_0$. As the angle between the two aromatic rings is increased further to 52° in 4,4'-diaminooctafluorobiphenyl (**6**), the conductance value also drops further, to $4.9 \times 10^{-4}$ $G_0$. When all four hydrogens on the proximal carbons of this molecule are replaced with methyls to give 2,6,2',6'-tetramethyl-4,4'-diaminobiphenyl (**8**) with a twist angle θ of about 88°, most counts in the histogram occur for low junction conductance values. In Figure 2C, we plot the peak conductance value measured for the seven molecules against $\cos^2\theta$ and fit the data with a $\cos^2\theta$ curve (dashed line in Figure 2C). The good fit indicates that the electronic effects of the substituents do not significantly alter the simple picture that junction conductance may be adjusted by simply decreasing the π-overlap between phenyl rings within the bridging molecule.

Further evidence for tunnelling conductance through these aromatic diamine junctions is provided by Figure 3A, which compares the histograms for a series of oligophenyls (molecules **1**, **4**, **9**) with 1, 2, and 3 phenyl rings. The inset shows that the plot of junction conductance versus molecule length fits an exponential form with a decay constant of 1.7 ± 0.1 per phenyl ring. This behaviour is direct evidence for non-resonant tunnelling transport through the amine-terminated molecules[15-19], with the measured decay constant being close to the estimate of 1.5 per phenyl ring that we obtain[20-22] for these molecules (see Supplementary Information). In Table 1, we also compare the measured conductance of all molecules with an estimated relative conductance, obtained by calculating the square of the tunnel coupling of each molecule and normalizing this value to the measured conductance of molecule **1**. This comparison indicates that the trend in the measured conductances is largely accounted for by the change in tunnelling with molecular conformation.

The results presented so far agree quite well with predictions of non-resonant tunnelling transport through static molecules. However, it is well established that the energy barriers for ring rotations in unsubstituted biphenyls such as **4** are about 0.1 eV or 2-3 kcal/mole[23] (the barriers are altered by the substituents in the molecules **5-8**, and substantially higher; see Supplementary Information). Therefore at room temperature, such molecules in the junction can be expected to fully explore low-energy rotations. The measurement times are long compared to the rate of molecular rotations in solution (each conductance step lasts a few milliseconds) therefore the conductance level measured at each step on a single trace is a thermal average over any fast dynamic fluctuations. But even though the rotation barrier is small for molecule **4**, our calculations indicate that thermal averaging still results in a conductance value that is close to the value characteristic of a static molecule in its lowest energy conformation.

The data in Figures 1-3 also reveal another interesting trend: as the molecule is granted additional rotational degrees of freedom, the conductance peak broadens. This is in contrast to the data for the diamino-alkane series[13], where the histogram peak widths are similar for all molecules and associated only with the modest variations in the electronic coupling through the Au-amine link. The peak broadening for the aromatics (widths in Table 1), is most clearly seen in Figure 3A for the oligophenyl series. The sharpest peak is observed for **1**, while the peaks in the histograms are systematically broader for biphenyl- and terphenyl-diamines (**4** and **9**), where the individual phenyl rings can rotate about the sigma bond that connects them.  Individual conductance traces show steps (Supplementary Information Figures S1 and S2) that occur over an expanded the range. We therefore hypothesize that in each junction formed during the measurements, the molecule is in a different average conformation enabled by its rotational degrees of freedom and that each different conformation has a different characteristic conductance. In solution, the thermally-averaged conformation for molecule **4** would have both the ring-ring twist angle and the two amine-ring twist angles close to their rotational barrier minima of 34° and 90° respectively (angles illustrated in Figure 3B). Upon junction formation, each Au-N link is pinned by the particular Au-N

6bonding environment. This provides a mechanism to skew the twist angles of the trapped molecule. The molecule responds by adjusting its three rotatable bonds (the C-C and the two C-N sigma bonds) to average twist values that differ from those in the unbound molecule, with the bond that has a lower energy barrier for rotation accommodating a larger change. This results in wider histograms, although the histogram peaks, which correspond to the most probable Au/molecule/Au conformations, are still located close to the lowest energy conformation.

**Supplementary Information** accompanies the paper on *Nature*'s website (http://www.nature.com).

**Acknowledgements** We thank Horst Stormer, Philip Kim and Julio Fernandez for useful discussions. This research was supported by the NSF Nanoscale Science and Engineering Center at Columbia University, New York State Office of Science (NYSTAR). CN thanks the Camille Dreyfus Teacher Scholar Program (2004) and the Alfred P. Sloan Fellowship Program (2004). J.E.K. thanks the American Chemical Society Division of Organic Chemistry for the graduate fellowship sponsored by Organic Syntheses. M.L.S. thanks the Material Research Science and Engineering Center program of the NSF.

Correspondence and requests for materials should be addressed L.V. (e-mail: latha@phys.columbia.edu) or M.S.H (msh2102@columbia.edu).

**Table 1: Molecule structure, measured conductances, calculated relative conductances, widths of the histogram peaks and the calculated twist angle (θ).**

| Molecule Number | Structure | Conductance ($G_0$) | | Peak Width* | Twist Angle |
|---|---|---|---|---|---|
| | | Measured | Calculated | | |
| 1 | 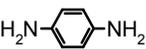 | $6.4 \times 10^{-3}$ | $6.4 \times 10^{-3}$ | 0.4 | — |
| 2 | 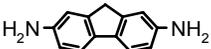 | $1.54 \times 10^{-3}$ | $2.1 \times 10^{-3}$ | 0.8 | 0 |
| 3 | 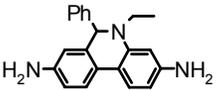 | $1.37 \times 10^{-3}$ | $2.2 \times 10^{-3}$ | 0.8 | 17 |
| 4 | 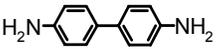 | $1.16 \times 10^{-3}$ | $1.6 \times 10^{-3}$ | 0.9 | 34 |
| 5 | 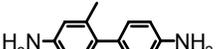 | $6.5 \times 10^{-4}$ | $1.2 \times 10^{-3}$ | 1.3 | 48 |
| 6 | 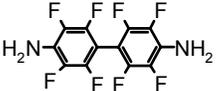 | $4.9 \times 10^{-4}$ | $7.1 \times 10^{-4}$ | 0.6 | 52 |
| 7 | 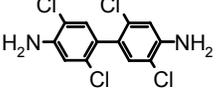 | $3.7 \times 10^{-4}$ | $5.8 \times 10^{-4}$ | 0.9 | 62 |
| 8 | 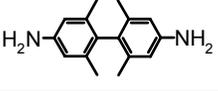 | $7.6 \times 10^{-5}$† | $6.4 \times 10^{-5}$ | NA† | 88 |
| 9 | 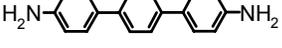 | $1.8 \times 10^{-4}$‡ | $3.5 \times 10^{-4}$ | 2.1 | — |

*Half widths at half maximum of the Lorentzian fit, normalized to the peak value.
†The histogram peak was determined after subtracting the Au histogram from the data as the raw data could not be fit with a Lorentzian hence a width could not be determined. ‡Determined from actual maximum of the raw data.

**Figure 1: Conductance measurements of 1,4-diaminobenzene and 2,7-diaminofluorene junctions** (A) Conductance histogram of **1** (green, constructed from 10000 traces and scaled by 10) and **2** (purple, constructed from 15000 and scaled by 15) on a log scale along with a control histogram without molecules (yellow, constructed from 6000 traces and scaled by 6). Peaks are clearly visible at $G_0$ and $2 \times G_0$ for all three curves and at $6.4 \times 10^{-3} G_0$ for **1** and at $1.5 \times 10^{-3} G_0$ for **2**. Also shown, is a histogram collected for an equimolar solution of both molecules (red curve, vertically offset, constructed from 4000 traces and scaled by 4). Inset: Lorentzian fits to the peak in histogram of **1**. The bin size is $10^{-5} G_0$ for all histograms and data were taken 25 mV. (B) Illustrations of a gap between a gold point-contact that can be bridged with either molecule **1** or **2** from the surrounding solution. (C) Sample conductance traces showing steps corresponding to the conductance of molecule **1** and **2**.

**Figure 2: Biphenyl junction conductance as a function of molecular twist angle** (A) Structures of a subset of the biphenyl series studied shown in order of increasing twist angle or decreasing conjugation. (B) Conductance histograms constructed for molecules **2** (purple, constructed from 15000 traces and scaled by 15), **4** (cyan, constructed from 7000 traces and scaled by 7), **6** (pink, constructed from 11000 traces and scaled by 11) and **8** (blue, constructed from 5000 traces and scaled by 5). Also shown is a control histogram without molecules (yellow, constructed from 6000 traces and scaled by 6). Arrows point to the peaks obtained from Lorentzian fits (solid black curves). All data were taken at 25 mV. (C) Position of the peaks for all the molecules studied plotted against $\cos^2(\theta)$, where $\theta$, the calculated twist angle for each molecule, is listed in Table 1. Error bars are determined from the standard deviation of the peak

locations determined from the fits to histograms of 1000 traces (see Supplementary Information.)

**Figure 3: Polyphenyl junction conductance as a function of the number of phenyl units** (A) Conductance histograms of **1** (green, constructed from 10000 traces and scaled by 10), **4** (cyan, constructed from 7000 traces and scaled by 7), and **9** (red, constructed from 3000 traces and scaled by 3) along with a control histogram without molecules (yellow, constructed from 6000 traces and scaled by 6). Bin size is $10^{-5}G_0$ for green trace (and is therefore offset vertically by a factor of 10) and $10^{-6}G_0$ for the other traces. Arrows point to the peaks. Lorentzian fits (solid black curves) are also shown. All data were taken at 25 mV. Inset: Exponential fit to peaks values shown on a semi-log scale. (B) Illustration of molecule **4** showing the three twist angles defined as the angle between the adjacent planes shown in the figure. In solution, the thermally averaged conformation has a ring-ring twist angle and amine-ring twist angle close to the potential minima of 34 and 90 degrees respectively.

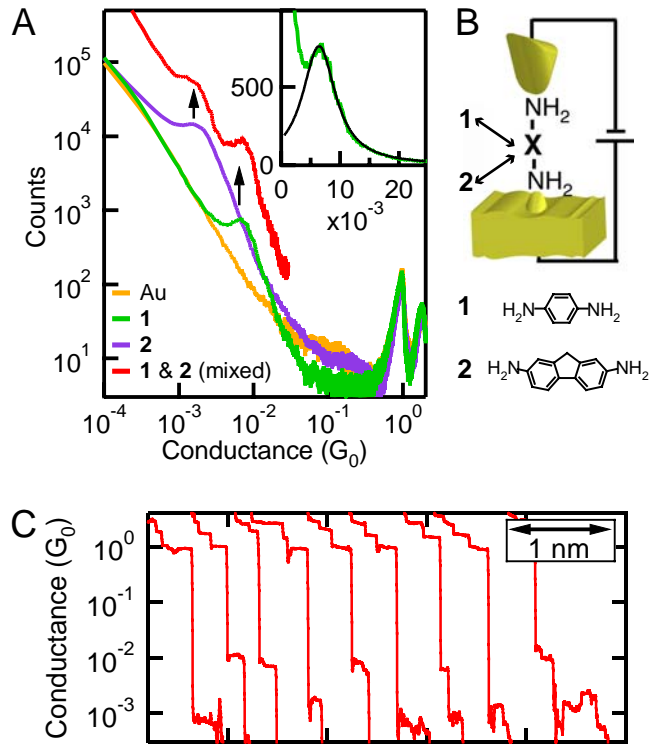

Figure 2

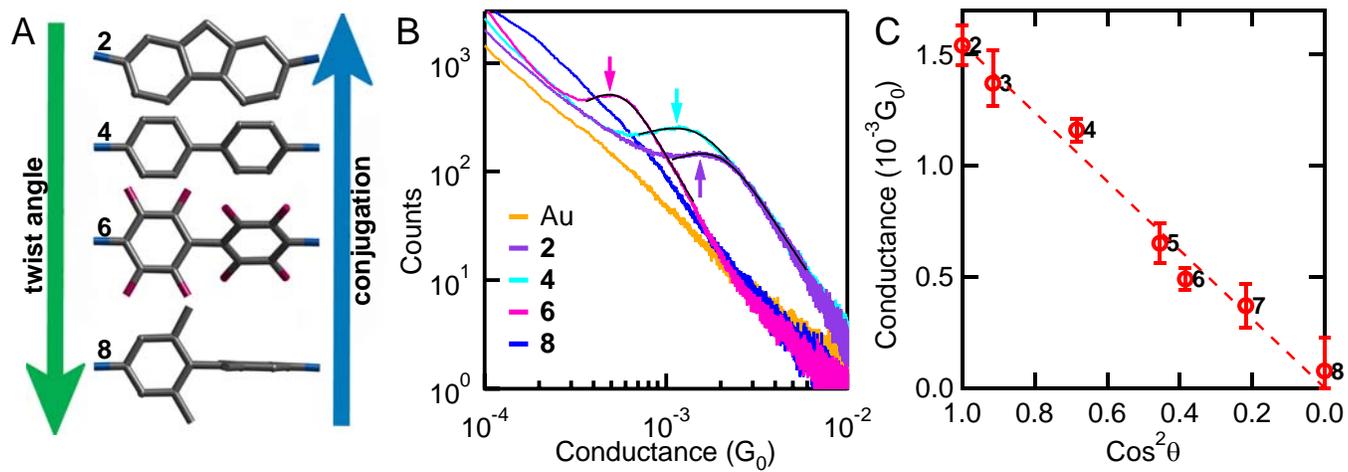

Figure 3

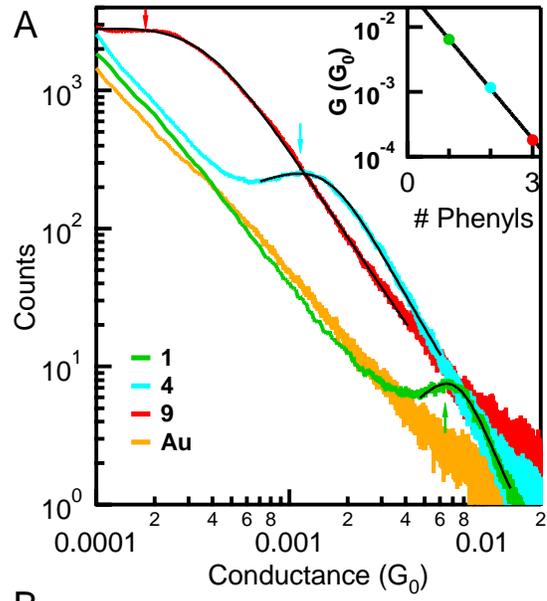

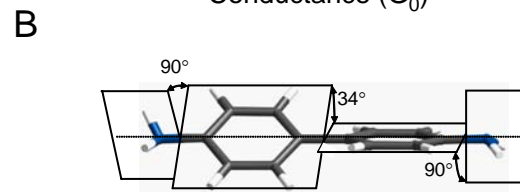